\documentclass[english,aps, longbibliography]{revtex4-2}
\usepackage[LGR,T1]{fontenc}
\usepackage[latin9]{inputenc}
\usepackage{amsmath}
\usepackage{amssymb}
\usepackage{graphicx}

\makeatletter

\DeclareRobustCommand{\greektext}{%
  \fontencoding{LGR}\selectfont\def\encodingdefault{LGR}}
\DeclareRobustCommand{\textgreek}[1]{\leavevmode{\greektext #1}}
\ProvideTextCommand{\~}{LGR}[1]{\char126#1}

\makeatother

\usepackage{babel}
\begin{document}
\title{Stochastic Thermodynamics of Cooperative Biomolecular Machines: Fluctuation
Relations and Hidden Detailed Balance Breaking}
\author{D. Evan Piephoff}
\author{Jianshu Cao}
\email{jianshu@mit.edu}

\affiliation{Department of Chemistry, Massachusetts Institute of Technology, Cambridge,
Massachusetts 02139, United States}
\begin{abstract}
We examine a biomolecular machine involving a driven, observable process
coupled to a hidden process in a kinetically cooperative manner. A
stochastic thermodynamics framework is employed to analyze a fluctuation
theorem for the first-passage time of the observable process under
nonequilibrium steady-state conditions. Based on a generic kinetic
model, we demonstrate that, along first-passage trajectories, entropy
production remains constant when the changes in stochastic entropy
and free energy of the machine are balanced, which corresponds to
zero net hidden flux through the initial state manifold. Under this
condition, which we define quite generally, this first-passage time
fluctuation theorem can be established, with its violation serving
as an experimentally detectable signature of hidden detailed balance
breaking (which we subsequently characterize). In addition, using
an enzymatic model, we show that the violation of our first-passage
time fluctuation theorem can be thought of as a consequence of the
breakdown of local detailed balance in the steps linking coarse-grained
states that correspond to the initial and intermediate state manifolds.
In the absence of hidden current, the fluctuation theorem is restored,
and a mesoscopic local detailed balance condition can be established,
which has implications for the thermodynamic analysis of driven, coarse-grained
systems. This work sheds significant light on the unique connections
between stochastic thermodynamic quantities and kinetic measurements
in complex cooperative networks.
\end{abstract}
\maketitle

\section{Introduction}

Recent advances in spectroscopic experimental techniques have provided
the ability to observe real-time trajectories of biomolecules at the
single-molecule level \citep{Moerner_2003,Park2007}. Such time traces
provide insights into microscopic mechanisms that are typically inaccessible
from ensemble-averaged measurements \citep{English2005}. In single-molecule
experiments, it is common to measure probability distribution functions
(PDFs) of the waiting times between detectable molecular events, such
as the first-passage time (i.e., the process completion time) PDF.

Biomolecular machines, including enzymes \citep{Seifert2011,Seifert_2012,Lu_Z_2023}
and motor proteins \citep{Svoboda1994,Keller2000,Sivak_2022}, consume
energy and dissipate heat to perform a given cellular function (e.g.,
catalysis, cargo transport, etc.). Accordingly, they operate out of
equilibrium, frequently in a nonequilibrium steady-state (NESS). In
the nonequilibrium scenario, a fluctuation theorem can demonstrate
properties of the PDF of a particular thermodynamic quantity (e.g.,
entropy production) \citep{Seifert_2012}. A time-based fluctuation
theorem was recently derived \citep{Roldan_2015,Neri_2017} for the
first-passage time of entropy production, that is, the time necessary
to produce a given amount of entropy. This relation implies equivalence
between the normalized forward and backward first-passage time PDFs
for entropy production. Based on chemical kinetics, this equivalence---referred
to as the generalized Haldane relation---has been demonstrated elsewhere
\citep{Qian_2006,Ge_2008} for the forward and backward first-passage
time PDFs for a generalized, one-dimensional (1D) enzymatic chain.

In this kinetic chain, the entropy produced along first-passage trajectories
is constant. However, for biomolecular machines involving a driven,
observable process coupled to a hidden process in a kinetically cooperative
\citep{Fersht_1985,Cao2011,Wu2012,Piephoff_2017,Mu_2021} manner,
this is not necessarily the case since such trajectories can start
and end in different underlying states; thus, this fluctuation theorem
no longer applies for the observable process first-passage time. Moreover,
single-molecule spectroscopic experiments have shown that slow, hidden
conformational fluctuations can occur on time scales commensurate
to those for the observable process \citep{English2005}; however,
many theoretical studies have neglected their role because the calculations
involved are quite complex.

We recently analyzed \citep{fpt-fluct-thm-arxiv} such a first-passage
time fluctuation theorem using the canonical model for this type of
system, a kinetic scheme for conformation-modulated single-enzyme
catalysis under NESS conditions (which has experimental relevance
to \textgreek{b}-galactosidase \citep{English2005} and human glucokinase
\citep{Mu_2021}). Our kinetic analysis revealed that in the absence
of hidden current, a fluctuation theorem can be established for the
observable process first-passage time, and we demonstrated that this
dramatic reduction is a general feature applicable to a wide variety
of cooperative biomolecular networks. The validity of this expression
can be tested experimentally, and its violation serves as a unique
signature of hidden detailed balance breaking. In addition, we characterized
the deviation from this relation, identifying a thermodynamic bound
on the kinetic branching ratio (a measure of directionality defined
as the ratio of the forward observable process probability to the
backward one).

Here, we employ a stochastic thermodynamics framework to interpret
our first-passage time fluctuation theorem results \citep{fpt-fluct-thm-arxiv}
and generalize them to a broad class of systems. The key results of
this paper are represented in eqs \ref{eq:deltaS-deltaF-rel}--\ref{eq:fluc-thm-tau},
\ref{eq:ldb_avgs}, and \ref{eq:fluct-thm-avg-rates}. Based on a
generic kinetic model for a cooperative biomolecular machine, we demonstrate
that, along first-passage trajectories, entropy production remains
constant when the changes in stochastic entropy and free energy of
the machine are equal, which corresponds to zero net hidden flux through
the initial state manifold (i.e., hidden detailed balance). Under
this condition, which we define quite generally, our first-passage
time fluctuation theorem is restored. Additionally, using an enzymatic
model, we show that the violation of this relation can be explained
as a consequence of the breakdown of local detailed balance in the
steps linking coarse-grained states corresponding to the initial and
intermediate state manifolds. In the absence of hidden current, the
fluctuation theorem is restored, and a mesoscopic local detailed balance
condition can be established, which has implications for the analysis
of coarse-grained systems. Lastly, we characterize the deviation from
hidden equilibrium by analyzing the kinetic branching ratio for a
more complex model, recovering our previously derived thermodynamic
bound.

\section{Generic Kinetic Model}

\begin{figure}
(a)\includegraphics[width=3in]{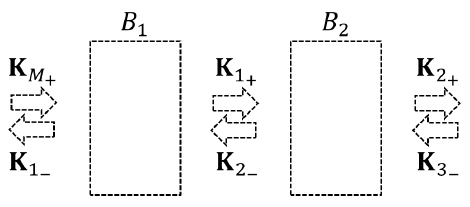}

(b)\includegraphics[width=3.5in]{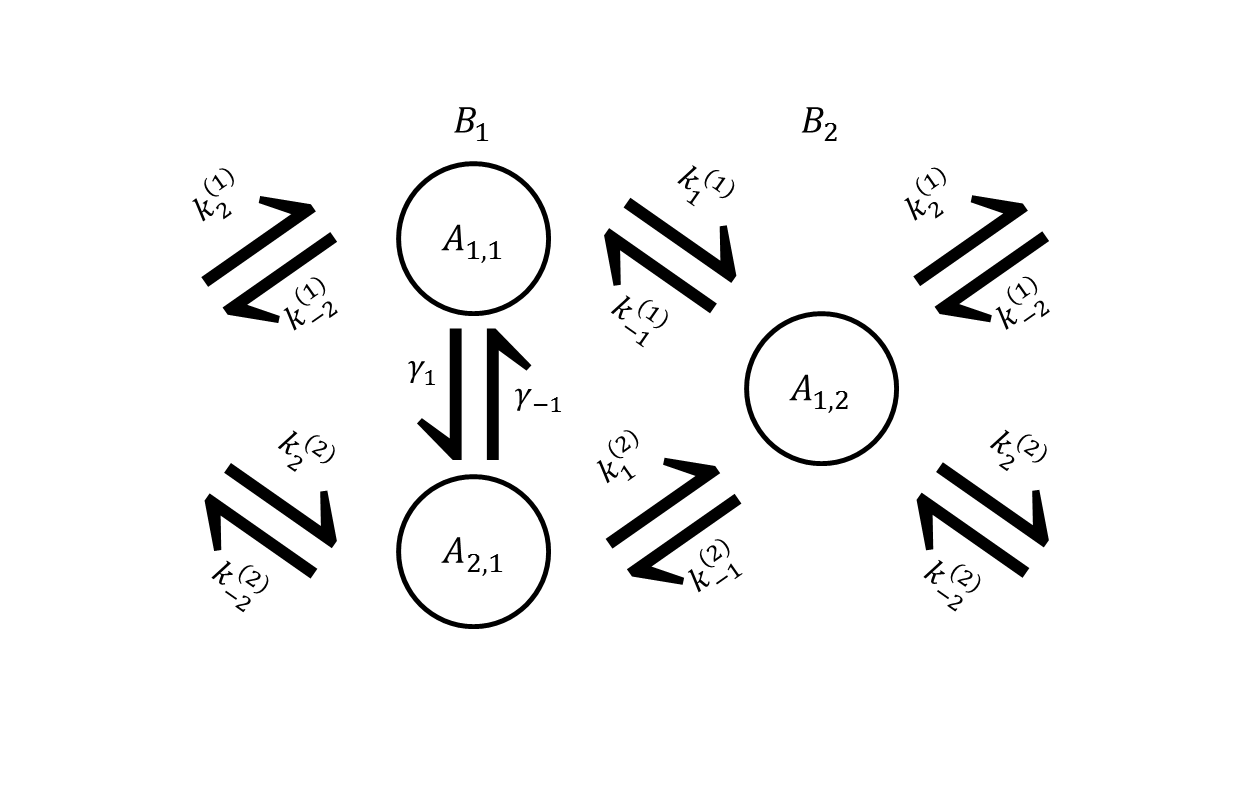}(c)\includegraphics[width=2.5in]{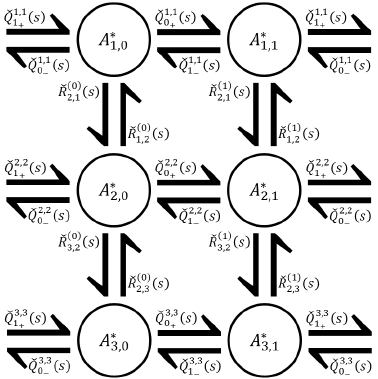}

\caption{\label{fig:gen-model}(a) Generic model for a biomolecular machine
with kinetic cooperativity under NESS conditions. The machine undergoes
a driven, observable, cyclic process cooperatively coupled to a hidden
process with dynamics occurring within the state manifolds, $\left\{ B_{m}\right\} $,
which have largely arbitrary internal topologies and connectivities
(see text for details). Transitions between manifolds are designated
here as $\left\{ \mathbf{K}_{\pm m}\right\} $. (b)--(c) Examples
of underlying schemes corresponding to the generic kinetic model in
(a). The rates of the transitions between the discrete states are
represented by $\left\{ k_{\pm m}^{k,l}\right\} $ and $\left\{ \gamma_{k,l}^{\left(m\right)}\right\} $.
}
\end{figure}

We begin by considering the generic model in Figure \ref{fig:gen-model}a
for a biomolecular machine with kinetic cooperativity \citep{Fersht_1985,Cao2011,Wu2012,Piephoff_2017}.
Examples of biomolecular machines include molecular motors as well
as single enzymes catalyzing the conversion of a substrate to a product.
Here, the machine undergoes a driven process (e.g., enzyme turnover)
cyclic about (i.e., it begins and ends in) $B_{1}$, the initial state
manifold. The overall process is reversible, with both the forward
and backward processes traversing the state manifold $B_{2}$. The
occurrence of a forward (rightward) $B_{2}$-to-$B_{1}$ kinetic manifold
transition, directly following a $B_{1}$-to-$B_{2}$ forward transition,
marks the completion of the forward process, while the occurrence
of a backward (leftward) $B_{2}$-to-$B_{1}$ manifold transition,
directly following a $B_{1}$-to-$B_{2}$ backward transition, marks
the completion of the backward process. We consider the basic direction
of propagation between state manifolds to be the observable direction
of the system, with the corresponding driven process referred to as
the observable process.

The observable process is cooperatively coupled to a hidden process
(or potentially a set of hidden processes) having dynamics occurring
within the manifolds, that of which is the result of simple thermal
changes not due to an external driving source (e.g., conformational
changes). Each state in $B_{1}$ and $B_{2}$ has an observable transition
entering and exiting it; otherwise, the internal topologies of the
manifolds are arbitrary, along with their connectivities. The underlying
kinetic scheme (see Figures \ref{fig:gen-model}b and \ref{fig:gen-model}c
for examples) is a network of discrete states with Markovian transition
dynamics \footnote{For thermodynamic consistency, it is required that all kinetic steps
be reversible.}. The machine is embedded in an aqueous solution that serves as a
heat bath with a constant temperature $T$. Driving sources are taken
to be time-independent; therefore, in the long-time limit, the system
reaches a NESS.

For our model in Figure \ref{fig:gen-model}a, we define a state coordinate
$x$ for the machine such that $x\in\mathbb{N}$. The network is described
by $A\left(x\right)$, where the state of the machine corresponding
to $x=i$ is represented as $A\left(x=i\right)=A_{i}$. We can subdivide
$x$ into two separate coordinates, $x_{1}$ and $x_{2}$, where $x_{1}$
corresponds to the hidden direction, and $x_{2}$ corresponds to the
observable one. Now, the network can be described by $A\left(x_{1},x_{2}\right)$,
with the state corresponding to $x_{1}=l,x_{2}=m$ represented as
$A\left(x_{1}=l,x_{2}=m\right)=A_{l,m}$. We can envisage the scheme
as a network of manifolds described by $B\left(x_{2}\right)$, where
the manifold of states corresponding to $x_{2}=m$ is represented
as $B\left(x_{2}=m\right)=B_{m}$. The rate of the hidden transition
from $A_{l,m}$ to $A_{k,m}$ is represented as $\gamma_{k,l}^{\left(m\right)}$;
the rate of the forward (backward) observable transition from $A_{l,1}$
to $A_{k,2}$ ($A_{l,2}$ to $A_{k,1}$) is $k_{1}^{k,l}$ ($k_{-1}^{k,l}$);
and the rate of the forward (backward) periodic observable transition
from $A_{l,2}$ to $A_{k,1}$ ($A_{l,1}$ to $A_{k,2}$) is $k_{2}^{k,l}$
($k_{-2}^{k,l}$).

A trajectory of duration $t$ is represented as $x\left(t\right)$.
Changes along $x\left(t\right)$ are defined as proceeding from $t=0$
to $t$, with the total entropy production along it represented by
$\Delta S^{\mathrm{tot}}\left[x\left(t\right)\right]$. We define
$\tau_{\pm}$ as the forward/backward first-passage time for the observable
process, i.e., the time necessary to complete an iteration of the
forward/backward process, while avoiding the completion of the backward/forward
one. An individual trajectory corresponding to such an iteration
is referred to as a forward/backward first-passage trajectory and
is designated as $x_{\pm}\left(\tau_{\pm}\right)$. The work applied
along such a trajectory, $\pm w$, which may include chemical work
(see Section III for an explanation), is referred to as the forward/backward
first-passage work, with $\Delta s^{\mathrm{tot}}=w/T$. That is,
we define $\Delta s^{\mathrm{tot}}$ as the entropy production associated
with the first-passage work. The unnormalized PDF of $\tau_{\pm}$
is represented as $P_{\pm}\left(\tau_{\pm}\right)$. When all forward/backward
first-passage trajectories produce entropy $\pm\Delta s^{\mathrm{tot}}$,
we can write a fluctuation theorem for $\tau_{\pm}$ that relates
the ratio $P_{+}\left(t\right)/P_{-}\left(t\right)$ exponentially
to $\Delta s^{\mathrm{tot}}$ \citep{Qian_2006,Roldan_2015,Neri_2017},
which we refer to as the first-passage time fluctuation theorem (see
eq \ref{eq:fluc-thm-tau} below). However, in a kinetically cooperative
biomolecular machine, the entropy produced along first-passage trajectories
is no longer constant, since they may begin and end in different underlying
states (as shown in Figure \ref{fig:enz-model}b below), resulting
in a breakdown of this relation. In order to analyze this first-passage
time fluctuation theorem in the context of our generic model in Figure
\ref{fig:gen-model}a, we will employ a stochastic thermodynamics
framework to examine $\Delta S^{\mathrm{tot}}\left[x_{\pm}\left(\tau_{\pm}\right)\right]$.

\section{Entropy Production Along Fluctuating Trajectories}

Here, we make some basic definitions for thermodynamic quantities
as they pertain to the system described above. We can write an entropy
balance along $x\left(t\right)$ as $\Delta S^{\mathrm{tot}}\left[x\left(t\right)\right]=\Delta S^{\mathrm{sto}}\left[x\left(t\right)\right]+\Delta S^{\mathrm{mach}}\left[x\left(t\right)\right]+\Delta S^{\mathrm{med}}\left[x\left(t\right)\right]$,
with the entropy change of the surrounding medium given by $\Delta S^{\mathrm{med}}\left[x\left(t\right)\right]=-Q\left[x\left(t\right)\right]/T$,
where $-Q\left[x\left(t\right)\right]$ is the dissipated heat. For
convenience, we have incorporated the entropic contribution of the
solution into $Q\left[x\left(t\right)\right]$ (see ref \citep{Seifert_2012}
for further details). The change in stochastic entropy of the system
along $x\left(t\right)$ is expressed as

\begin{equation}
\Delta S^{\mathrm{sto}}\left[x\left(t\right)\right]=-k_{\mathrm{B}}\ln\left(\frac{\rho^{\mathrm{s}}\left[x\left(t\right)\right]}{\rho^{\mathrm{s}}\left[x\left(0\right)\right]}\right)\label{eq:deltaSsto-defn}
\end{equation}
where $\rho^{\mathrm{s}}\left(x\right)$ is the stationary population
distribution, with $\rho^{\mathrm{s}}\left(x=i\right)=\rho_{i}^{\mathrm{s}}$,
and $k_{\mathrm{B}}$ is the Boltzmann constant. The energy of $A_{i}$
is represented as $E_{i}^{\mathrm{mach}}$, with corresponding intrinsic
entropy $S_{i}^{\mathrm{mach}}$ and free energy $F_{i}^{\mathrm{mach}}=E_{i}^{\mathrm{mach}}-TS_{i}^{\mathrm{mach}}$
\footnote{For an extended discussion of intrinsic entropy, we refer readers
to Seifert's work in refs \citealp{Seifert2011} and \citealp{Seifert_2012}.}. We can write a first law-like energy balance along $x\left(t\right)$
as $Q\left[x\left(t\right)\right]+W\left[x\left(t\right)\right]=\Delta E^{\mathrm{mach}}\left[x\left(t\right)\right]=\Delta F^{\mathrm{mach}}\left[x\left(t\right)\right]+T\Delta S^{\mathrm{mach}}\left[x\left(t\right)\right]$,
where $W\left[x\left(t\right)\right]$ is the work applied to the
machine (in the observable direction). In accordance with our above
definition of heat, $W\left[x\left(t\right)\right]$ may include chemical
work, i.e., the negative of the free energy change of the solution
due to a reaction with stoichiometrically differing total chemical
potentials between the reactants and products \citep{Seifert_2012}
(example in Section V below). In our system, $W\left[x_{\pm}\left(\tau_{\pm}\right)\right]=\pm w$;
that is, the forward/backward first-passage work is path-independent.
The entropy balance above can then be rewritten as

\begin{equation}
\Delta S^{\mathrm{tot}}\left[x\left(t\right)\right]=\Delta S^{\mathrm{sto}}\left[x\left(t\right)\right]+\left(W\left[x\left(t\right)\right]-\Delta F^{\mathrm{mach}}\left[x\left(t\right)\right]\right)/T\label{eq:S_bal_W_deltaF}
\end{equation}

We now connect $\Delta F^{\mathrm{mach}}$ and $W$ to the rates of
individual transitions, such as those in the examples depicted in
Figures \ref{fig:gen-model}b and \ref{fig:gen-model}c. We define
a discrete transition coordinate $z$, where the free energy change
and applied work associated with the transition corresponding to $z=\xi$
are represented as $\Delta F^{\mathrm{mach}}\left(z=\xi\right)=\Delta_{\xi}F^{\mathrm{mach}}$
and $W\left(z=\xi\right)=W_{\xi}$, respectively. It is noted that
an individual transition (in the observable direction) can involve
multiple types of work (e.g., chemical, mechanical, etc.), such that
$W_{\xi}=\sum_{\eta}W_{\xi_{\eta}}$, where the $\eta$ subscript
corresponds to a particular type of work \footnote{We note that for a mechanically driven system, such as a molecular
motor, under a force $f$ with a distance step size $d_{\xi}$, $W_{\xi}=fd_{\xi}$
(in addition to any chemical or other work contributions).}. The coordinate $z_{x\left(t\right)}\left(\tau_{j}\right)$ corresponds
to the transition completed at $\tau_{j}$, the time at which the
$j$-th transition (which may or may not be a repeated transition)
is completed along $x\left(t\right)$. Along such a trajectory, $\Delta F^{\mathrm{mach}}\left[x\left(t\right)\right]=\sum_{j}\Delta F^{\mathrm{mach}}\left[z_{x\left(t\right)}\left(\tau_{j}\right)\right]$
and $W\left[x\left(t\right)\right]=\sum_{j}W\left[z_{x\left(t\right)}\left(\tau_{j}\right)\right]$.
For observable transition $\lambda$ with rate $k_{\lambda}$ and
hidden transition $\mu$ with rate $\gamma_{\mu}$, local detailed
balance \citep{Seifert2011,Seifert_2012} requires that

\begin{equation}
\frac{k_{\lambda}}{k_{\lambda^{\dagger}}}=\exp\left[-\beta\left(\Delta_{\lambda}F^{\mathrm{mach}}-W_{\lambda}\right)\right]\label{eq:ldb-k}
\end{equation}

\begin{equation}
\frac{\gamma_{\mu}}{\gamma_{\mu^{\dagger}}}=\exp\left[-\beta\Delta_{\mu}F^{\mathrm{mach}}\right]\label{eq:ldb-g}
\end{equation}
where the $\dagger$ superscript denotes the reverse transition, with
$\Delta_{\xi^{\dagger}}F^{\mathrm{mach}}=-\Delta_{\xi}F^{\mathrm{mach}}$
and $W_{\lambda^{\dagger}}=-W_{\lambda}$; $\beta=\left[k_{\mathrm{B}}T\right]^{-1}$.

\section{Analysis of First-Passage Trajectories}

Now, we examine entropy production for the generic model in Figure
\ref{fig:gen-model}a. Here, forward/backward first-passage trajectories
that begin and end in the same underlying state (within $B_{1}$)
produce entropy $\pm\Delta s^{\mathrm{tot}}$. However, because such
trajectories can begin and end in different states (as shown in Figure
\ref{fig:enz-model}b below), $\Delta S^{\mathrm{tot}}\left[x_{\pm}\left(\tau_{\pm}\right)\right]$
is not necessarily constant, as $\Delta S^{\mathrm{sto}}\left[x_{\pm}\left(\tau_{\pm}\right)\right]$
and $\Delta F^{\mathrm{mach}}\left[x_{\pm}\left(\tau_{\pm}\right)\right]$
can be nonzero. In order to achieve $\Delta S^{\mathrm{tot}}\left[x_{\pm}\left(\tau_{\pm}\right)\right]=\pm\Delta s^{\mathrm{tot}}$,
it is required that (see eq \ref{eq:S_bal_W_deltaF})

\begin{equation}
\Delta S^{\mathrm{sto}}\left[x_{\pm}\left(\tau_{\pm}\right)\right]=\Delta F^{\mathrm{mach}}\left[x_{\pm}\left(\tau_{\pm}\right)\right]/T\label{eq:deltaS-deltaF-rel}
\end{equation}
i.e., that the changes in stochastic entropy and free energy of the
machine be balanced along first-passage trajectories. The $A_{l,m}$
stationary population is represented as $\rho^{\mathrm{s}}\left(x_{1}=l,x_{2}=m\right)=\rho_{l,m}^{\mathrm{s}}$.
From eqs \ref{eq:deltaSsto-defn}, \ref{eq:ldb-k}, and \ref{eq:ldb-g},
we find that eq \ref{eq:deltaS-deltaF-rel} is satisfied under the
hidden detailed balance condition,

\begin{equation}
\frac{\rho_{k,1}^{\mathrm{s}}}{\rho_{l,1}^{\mathrm{s}}}=\frac{\gamma_{k,l}^{\left(1\right)}}{\gamma_{l,k}^{\left(1\right)}}\qquad\forall k,l\label{eq:hid-db}
\end{equation}
which corresponds to

\begin{equation}
J_{k,l}=0\qquad\forall k,l
\end{equation}
where the stationary hidden population flux from $A_{l,1}$ to $A_{k,1}$
is given by $J_{k,l}=\gamma_{k,l}^{\left(1\right)}\rho_{l,1}^{\mathrm{s}}-\gamma_{l,k}^{\left(1\right)}\rho_{k,1}^{\mathrm{s}}$.
It is noted that the constraints of local detailed balance (eqs \ref{eq:ldb-k}
and \ref{eq:ldb-g}) alone are generally insufficient in satisfying
hidden detailed balance, as demonstrated in eqs \ref{eq:ldb-enz-k}--\ref{eq:hdb-enz}
below. Equation \ref{eq:hid-db} is a stricter condition than eqs
\ref{eq:ldb-k} and \ref{eq:ldb-g}, and it corresponds to zero net
hidden flux (i.e., current) through the initial state manifold.

Under eq \ref{eq:hid-db}, all forward/backward first-passage trajectories
now produce entropy $\pm\Delta s^{\mathrm{tot}}$, so for the generic
model in Figure \ref{fig:gen-model}a, we can write the first-passage
time fluctuation theorem \footnote{Note that the time dependence on the left-hand side of eq \ref{eq:fluc-thm-tau}
divides out.} (consistent with our previously reported kinetic analysis \citep{fpt-fluct-thm-arxiv})

\begin{equation}
\frac{P_{+}\left(t\right)}{P_{-}\left(t\right)}=\exp\left[\frac{\Delta s^{\mathrm{tot}}}{k_{\mathrm{B}}}\right]\label{eq:fluc-thm-tau}
\end{equation}
recovering the form obtained elsewhere for a 1D kinetic chain \citep{Qian_2006}.
That is, in the absence of hidden current, $P_{+}\left(t\right)/P_{-}\left(t\right)$
dramatically reduces to a fluctuation theorem equivalent to the one
derived by Roldán and Neri et al. \citep{Roldan_2015,Neri_2017} for
the first-passage time of entropy production. This result is surprising
because, under eq \ref{eq:hid-db}, $P_{\pm}\left(t\right)$ does
not generally reduce to the 1D chain form, but the ratio $P_{+}\left(t\right)/P_{-}\left(t\right)$
does, indicating a unique reduction in how the forward and backward
observable processes relate to each another. The normalized PDF corresponding
to $P_{\pm}\left(t\right)$ is given by $\phi_{\pm}\left(t\right)=P_{\pm}\left(t\right)/p_{\pm}$,
with forward/backward observable process probability $p_{\pm}=\int_{0}^{\infty}dtP_{\pm}\left(t\right)$,
where $p_{+}+p_{-}=1$. Equation \ref{eq:fluc-thm-tau} implies the
symmetry relation $\phi_{+}\left(t\right)=\phi_{-}\left(t\right)$,
which is referred to as the generalized Haldane relation and has been
derived for a 1D enzymatic chain reaction \citep{Qian_2006,Ge_2008}.
Similarly, it was shown that for a 1D kinetic chain involving a motor
protein, the mean forward and backward first-passage times are equal
\citep{Kolomeisky2005}. The PDF $\phi_{\pm}\left(t\right)$ can be
experimentally measured; therefore, the generalized Haldane relation
can be tested, and its violation---which implies a violation of the
first-passage time fluctuation theorem (eq \ref{eq:fluc-thm-tau})---serves
as a unique signature of hidden detailed balance breaking. It is noted
that $P_{\pm}\left(t\right)$ (as well as $w$) can typically also
be measured; thus, the first-passage time fluctuation theorem can
be directly tested itself. We recently proved this signature using
a novel technique to analyze the kinetics; the analysis presented
herein interprets this result through the lens of stochastic thermodynamics
and generalizes it to a broad class of driven processes with coupled,
hidden dynamics.

\section{Application to Single-Enzyme Catalysis}

\begin{figure}
(a)\includegraphics[width=4in]{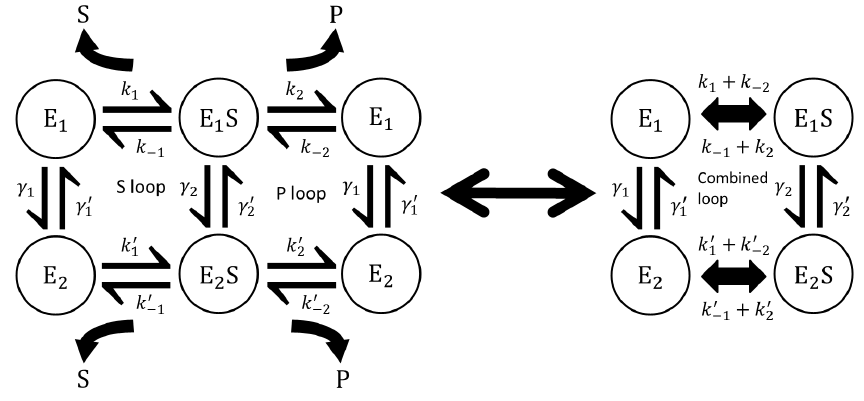}

(b)\includegraphics[width=3in]{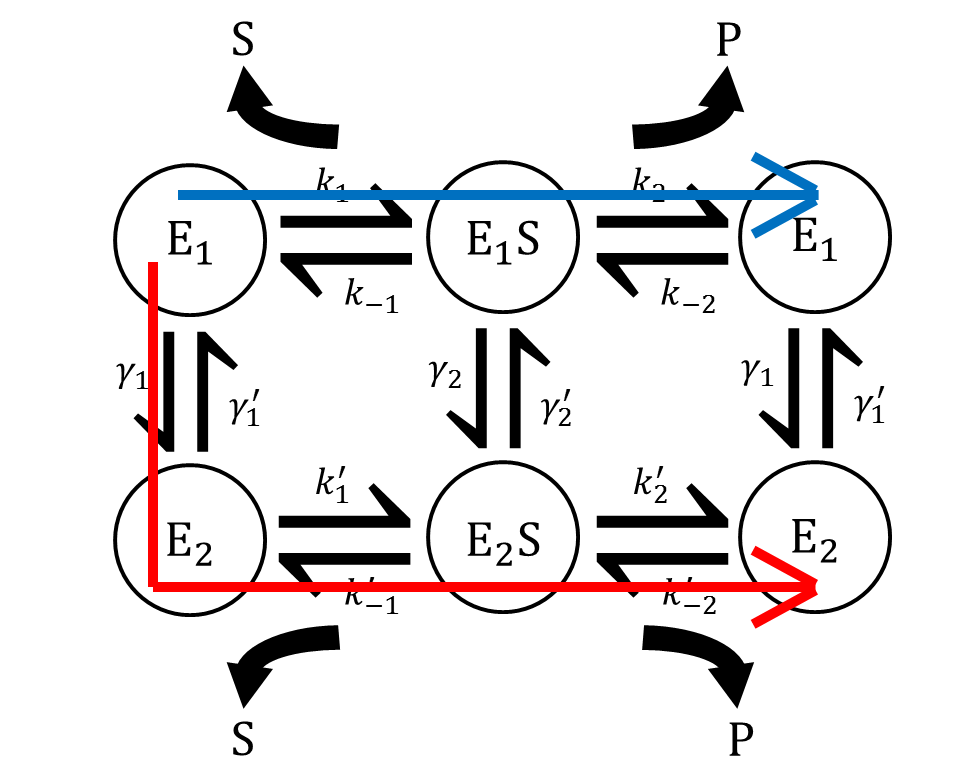}

\caption{\label{fig:enz-model}(a) Adaptation of the generic kinetic model
in Figure \ref{fig:gen-model}a to single-enzyme turnover with conformational
interconversion. The enzyme reversibly catalyzes the conversion of
a substrate to a product (see text for details). The reaction is cooperatively
coupled to a hidden process, with the enzyme undergoing slow conformational
changes within $B_{1}$ (state manifold for the free enzyme) and $B_{2}$
(state manifold for the substrate-bound enzymatic complex). The right-hand
side is a representation of the scheme wherein the two steps in each
reaction pathway are folded onto each other, resulting in a conformational
loop with a corresponding population current $J$. (b) Depiction of
two first-passage trajectories for the model in (a), each producing
a different amount of entropy, with the blue one starting and ending
in the same underlying state, and the red one doing so in different
states.}
\end{figure}

To demonstrate the above results with an illustrative example, we
now adapt the generic kinetic model in Figure \ref{fig:gen-model}a
to an enzymatic reaction with conformational interconversion (shown
in Figure \ref{fig:enz-model}a). Here, a single enzyme reversibly
catalyzes the conversion of a substrate, $\mathrm{S}$, to a product,
$\mathrm{P}$. The free enzyme, $\mathrm{E}$ (state manifold represented
by $B_{1}$), can reversibly bind the substrate, resulting in the
formation of the substrate-bound enzymatic complex, $\mathrm{ES}$
(state manifold represented by $B_{2}$), which can then reversibly
undergo product formation. Substrate is consumed to form product in
the forward observable process, and product is consumed to form substrate
in the backward one. The single enzyme is embedded in a solution of
substrate and product, such that the substrate and product concentrations
(and chemical potentials) remain fixed here, and the nonlinear substrate
binding and reverse product formation kinetic transitions are treated
as pseudolinear. The reaction is cooperatively coupled to a hidden
process, as the enzyme undergoes slow conformational changes within
$B_{1}$ and $B_{2}$. Such conformation-modulated enzymatic models
have experimental relevance to \textgreek{b}-galactosidase \citep{English2005}
and human glucokinase \citep{Mu_2021}. For convenience, our notation
has been modified here, such that the $A_{l,1}$-to-$A_{l,2}$ ($A_{l,2}$-to-$A_{l,1}$)
transition rates are now represented as $k_{1}^{\left(l\right)}$
and $k_{-2}^{\left(l\right)}$ ($k_{2}^{\left(l\right)}$ and $k_{-1}^{\left(l\right)}$),
and the $A_{1,m}$-to-$A_{2,m}$ ($A_{2,m}$-to-$A_{1,m}$) transition
rate is now represented as $\gamma_{1}^{\left(m\right)}$ ($\gamma_{-1}^{\left(m\right)}$).

Let $\mu^{\mathrm{S}}$ represent the chemical potential of the substrate,
and $\mu^{\mathrm{P}}$ represent that of the product. The reaction
process is driven by the difference in chemical potential between
the substrate and product (i.e., the chemical affinity), $-\Delta\mu$,
that is, $w=\mu^{\mathrm{S}}-\mu^{\mathrm{P}}\equiv-\Delta\mu$. From
eqs \ref{eq:ldb-k} and \ref{eq:ldb-g}, it is seen that local detailed
balance constrains the transition rates here as \footnote{It is noted that eq \ref{eq:ldb-enz-conf} represents the local detailed
balance condition for the closed substrate loop. A similar condition
can be written for the product loop, $\gamma_{1}^{\left(1\right)}k_{-2}^{\left(2\right)}\gamma_{-1}^{\left(2\right)}k_{2}^{\left(1\right)}/\left(\gamma_{-1}^{\left(1\right)}k_{2}^{\left(2\right)}\gamma_{1}^{\left(2\right)}k_{-2}^{\left(1\right)}\right)=1$,
which is implied by eqs \ref{eq:ldb-enz-k} and \ref{eq:ldb-enz-conf};
however, only two independent constraints can be imposed.}

\begin{equation}
\frac{k_{1}^{\left(1\right)}k_{2}^{\left(1\right)}}{k_{-1}^{\left(1\right)}k_{-2}^{\left(1\right)}}=\frac{k_{1}^{\left(2\right)}k_{2}^{\left(2\right)}}{k_{-1}^{\left(2\right)}k_{-2}^{\left(2\right)}}=\exp\left[-\beta\Delta\mu\right]\label{eq:ldb-enz-k}
\end{equation}

\begin{equation}
\frac{\gamma_{1}^{\left(1\right)}k_{1}^{\left(2\right)}\gamma_{-1}^{\left(2\right)}k_{-1}^{\left(1\right)}}{\gamma_{-1}^{\left(1\right)}k_{-1}^{\left(2\right)}\gamma_{1}^{\left(2\right)}k_{1}^{\left(1\right)}}=1\label{eq:ldb-enz-conf}
\end{equation}
Therefore, the kinetics are described by ten independent rates. The
hidden (conformational) population current (depicted on the right-hand
side of Figure \ref{fig:enz-model}a) is given by $J=J_{2,1}=\rho_{1,1}^{\mathrm{s}}\gamma_{1}^{\left(1\right)}-\rho_{2,1}^{\mathrm{s}}\gamma_{-1}^{\left(1\right)}$.
In addition, $J/\gamma^{\left(1\right)}\propto\left[\gamma_{1}^{\left(1\right)}u_{1}^{\left(2\right)}\gamma_{-1}^{\left(2\right)}u_{-1}^{\left(1\right)}/\left(\gamma_{-1}^{\left(1\right)}u_{-1}^{\left(2\right)}\gamma_{1}^{\left(2\right)}u_{1}^{\left(1\right)}\right)-1\right]$,
where $u_{\pm1}^{\left(l\right)}=k_{\pm1}^{\left(l\right)}+k_{\mp2}^{\left(l\right)}$,
and $\gamma^{\left(1\right)}$ is a scale given by $\gamma^{\left(1\right)}=\gamma_{1}^{\left(1\right)}+\gamma_{-1}^{\left(1\right)}$.
The hidden (conformational) detailed balance condition, under which
$J=0$, can then be expressed here as

\begin{equation}
\frac{\gamma_{1}^{\left(1\right)}u_{1}^{\left(2\right)}\gamma_{-1}^{\left(2\right)}u_{-1}^{\left(1\right)}}{\gamma_{-1}^{\left(1\right)}u_{-1}^{\left(2\right)}\gamma_{1}^{\left(2\right)}u_{1}^{\left(1\right)}}=1\label{eq:hdb-enz}
\end{equation}
From eqs \ref{eq:ldb-enz-k}--\ref{eq:hdb-enz}, we see that local
detailed balance alone is insufficient in satisfying hidden detailed
balance. If the two steps in each reaction pathway were folded onto
each other (as shown on the right-hand side of Figure \ref{fig:enz-model}a),
then the satisfaction of hidden detailed balance would correspond
to the probability of traversing the resulting loop being directionally
invariant. When hidden detailed balance (eq \ref{eq:hdb-enz}) is
satisfied here, we can write the first-passage time fluctuation theorem
(consistent with our previously reported kinetic analysis)

\begin{equation}
\frac{P_{+}\left(t\right)}{P_{-}\left(t\right)}=\exp\left[-\beta\Delta\mu\right]\label{eq:tau-fluc-thm-deltamu}
\end{equation}
which implies the generalized Haldane relation (see Section IV).
The validity of eq \ref{eq:tau-fluc-thm-deltamu} can be experimentally
tested, with its violation serving as a unique signature of hidden
detailed balance breaking.

\section{Coarse-Graining and Mesoscopic Local Detailed Balance}

In the absence of slow hidden (i.e., conformational) dynamics, the
enzymatic model in Figure \ref{fig:enz-model}a is described by a
1D kinetic chain with rates $\left\{ k_{\pm m}\right\} $. In this
picture, the rates for each step obey local detailed balance, as they
are related by eq \ref{eq:ldb-k}. As a result, it can be readily
shown in this case that \citep{Qian_2006} $P_{+}\left(t\right)/P_{-}\left(t\right)=k_{1}k_{2}/\left(k_{-1}k_{-2}\right)=\exp\left[\beta w\right]$.

In the presence of slow hidden dynamics (i.e., for the model in Figure
\ref{fig:enz-model}a), the observable (i.e., chemical) states are
described by the manifolds $\left\{ B_{m}\right\} $. The manifold
$B_{m}$ can be coarse-grained, such that the coarse-grained state
$\bar{B}_{m}$ represents the corresponding hidden (i.e., conformational)
ensemble-averaged observable state with stationary population $\bar{\rho}_{m}^{\mathrm{s}}=\sum_{l}\rho_{l,m}^{\mathrm{s}}$.
As a result of the coarse-graining, local detailed balance breaks
down in the steps linking $\bar{B}_{1}$ and $\bar{B}_{2}$ (i.e.,
mesoscopic local detailed balance cannot be established). That is,
for observable step $m$ with underlying rates $\left\{ k_{\pm m}^{\left(l\right)}\right\} _{l}$,
it is found that, in general, $\left\langle k_{m}\right\rangle /\left\langle k_{-m}\right\rangle \neq\left\langle \exp\left[-\beta\left(\Delta_{m}F^{\mathrm{enz}}-W_{m}\right)\right]\right\rangle $,
where the corresponding average forward rate is given by $\left\langle k_{m}\right\rangle =\sum_{l}\rho_{l,m}^{\mathrm{s}}k_{m}^{\left(l\right)}/\bar{\rho}_{m}^{\mathrm{s}}$,
with $\left\langle \exp\left[-\beta\left(\Delta_{m}F^{\mathrm{enz}}-W_{m}\right)\right]\right\rangle =\sum_{l}\rho_{l,m}^{\mathrm{s}}\left(k_{m}^{\left(l\right)}/k_{-m}^{\left(l\right)}\right)/\bar{\rho}_{m}^{\mathrm{s}}$,
and the average backward rates are given by $\left\langle k_{-1}\right\rangle =\sum_{l}\rho_{l,2}^{\mathrm{s}}k_{-1}^{\left(l\right)}/\bar{\rho}_{2}^{\mathrm{s}}$
and $\left\langle k_{-2}\right\rangle =\sum_{l}\rho_{l,1}^{\mathrm{s}}k_{-2}^{\left(l\right)}/\bar{\rho}_{1}^{\mathrm{s}}$.
Accordingly, $P_{+}\left(t\right)/P_{-}\left(t\right)\neq\left\langle k_{1}\right\rangle \left\langle k_{2}\right\rangle /\left(\left\langle k_{-1}\right\rangle \left\langle k_{-2}\right\rangle \right)$
here as well. However, when hidden detailed balance is satisfied,
local detailed balance is restored in the steps linking $\bar{B}_{1}$
and $\bar{B}_{2}$. That is, under eq \ref{eq:hdb-enz}, for observable
step $m$, we can establish the mesoscopic local detailed balance
condition (derivations in the Supporting Information),

\begin{equation}
\frac{\left\langle k_{m}\right\rangle }{\left\langle k_{-m}\right\rangle }=\left\langle \exp\left[-\beta\left(\Delta_{m}F^{\mathrm{enz}}-W_{m}\right)\right]\right\rangle \label{eq:ldb_avgs}
\end{equation}
and we find that

\begin{equation}
\frac{P_{+}\left(t\right)}{P_{-}\left(t\right)}=\frac{\left\langle k_{1}\right\rangle \left\langle k_{2}\right\rangle }{\left\langle k_{-1}\right\rangle \left\langle k_{-2}\right\rangle }=\exp\left[\beta w\right]\label{eq:fluct-thm-avg-rates}
\end{equation}
recovering eq \ref{eq:tau-fluc-thm-deltamu}. Thus, for this model,
the violation of the first-passage time fluctuation theorem can be
thought of as a consequence of the breakdown of local detailed balance
in the steps linking $\bar{B}_{1}$ and $\bar{B}_{2}$, that of which
is restored in the absence of hidden current. The concept of mesoscopic
local detailed balance and its connection to coarse-grained systems
will be explored further in a future publication.

\section{Hidden Detailed Balance Breaking}

\begin{figure}
(a)\includegraphics[width=3in]{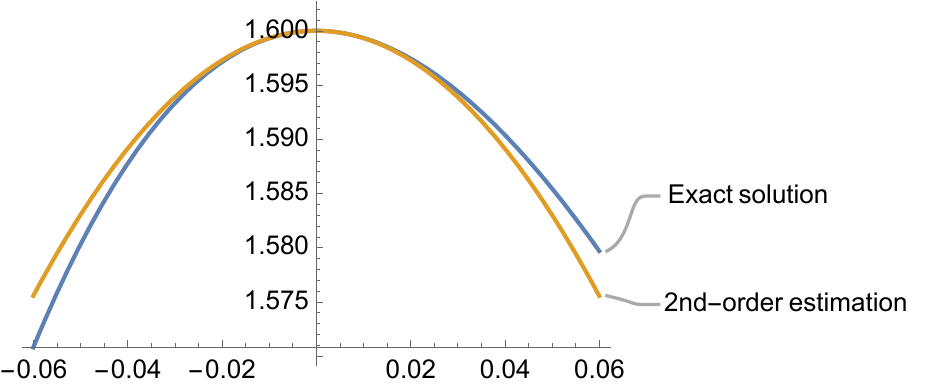}(b)\includegraphics[width=3in]{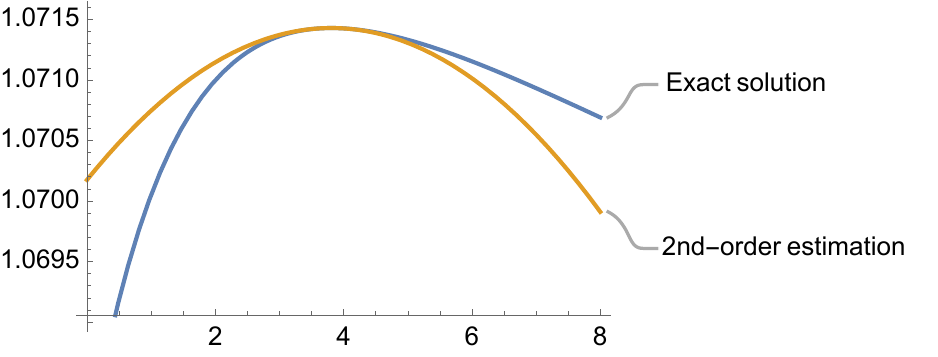}

(c)\includegraphics[width=2.5in]{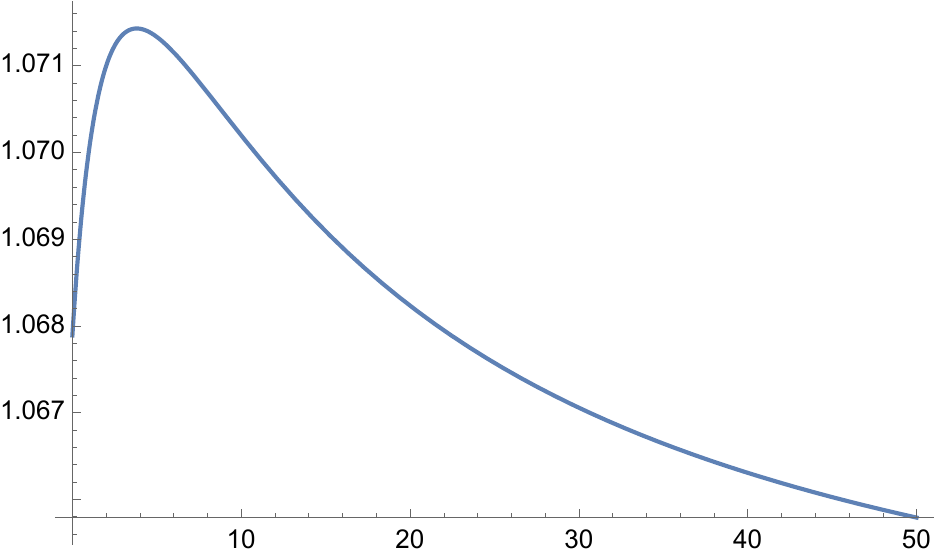}

\caption{\label{fig:pfpr-plots}Plots of $p_{+}/p_{-}$ against $J$ (a) and
$\gamma_{1}^{\left(1\right)}$ (b) and (c) for the model in Figure
\ref{fig:enz-model}a (see text for details). In (a) and (b), the
second-order estimation and the exact solution are shown; (c) shows
the full range of the exact solution in (b). In all panels, since
$w>0$, it is seen that the bound $p_{+}/p_{-}\protect\leq\exp\left[\beta w\right]$
is obeyed.}
\end{figure}

Now, we characterize hidden detailed balance breaking for the four-state
model in Figure \ref{fig:enz-model}a. We define $p_{+}/p_{-}$ as
the kinetic branching ratio for the observable process, which serves
as a measure of the deviation from hidden equilibrium and is obtainable
from experimental measurements. For $J\rightarrow0$ (i.e., near hidden
equilibrium), a second-order expansion of $\frac{p_{+}}{p_{-}}\left(J\right)$
about $J=0$ can be written as (see Supporting Information for further
details)

\begin{equation}
\frac{p_{+}}{p_{-}}\left(J\right)\approx\exp\left[\beta w\right]+aJ^{2}\label{eq:pfpr-expansion-J}
\end{equation}
where $a=\frac{1}{2}\frac{d^{2}}{dJ^{2}}\left[\frac{p_{+}}{p_{-}}\left(J\right)\right]_{J=0}$.
For $w>0$ ($<0$), $a<0$ ($>0$), i.e., the second-order estimation
of $\frac{p_{+}}{p_{-}}\left(J\right)$ has negative (positive) concavity,
consistent with the previously derived inequality $p_{+}/p_{-}\leq\exp\left[\beta w\right]$
($\geq\exp\left[\beta w\right]$).

Substituting in the local detailed balance constraints (eqs \ref{eq:ldb-enz-k}
and \ref{eq:ldb-enz-conf}) with $\gamma_{1}^{\left(1\right)}$ and
$k_{-2}^{\left(1\right)}$, and substituting in $J$ with $k_{2}^{\left(1\right)}$
(these substitution choices are arbitrary), we obtain $\exp\left[\beta w\right]=k_{1}^{\left(2\right)}k_{2}^{\left(2\right)}/\left(k_{-1}^{\left(2\right)}k_{-2}^{\left(2\right)}\right)$,
along with

\begin{equation}
\begin{split}aD & =\left(\gamma_{-1}^{\left(1\right)}\gamma_{-1}^{\left(2\right)}k_{-1}^{\left(1\right)}+\gamma_{-1}^{\left(1\right)}\gamma_{1}^{\left(2\right)}k_{-1}^{\left(2\right)}+\gamma_{-1}^{\left(1\right)}k_{-1}^{\left(1\right)}k_{-1}^{\left(2\right)}+\gamma_{-1}^{\left(2\right)}k_{-1}^{\left(1\right)}k_{1}^{\left(2\right)}+\gamma_{-1}^{\left(2\right)}k_{-1}^{\left(1\right)}k_{-2}^{\left(2\right)}+\gamma_{-1}^{\left(1\right)}k_{-1}^{\left(1\right)}k_{2}^{\left(2\right)}\right)\\
 & \,\,\,\,\,\,\,\times\left(\gamma_{1}^{\left(2\right)}k_{-1}^{\left(2\right)^{2}}k_{1}^{\left(1\right)}+\gamma_{-1}^{\left(2\right)}k_{-1}^{\left(1\right)}k_{-1}^{\left(2\right)}k_{1}^{\left(2\right)}+\gamma_{-1}^{\left(2\right)}k_{-1}^{\left(2\right)}k_{1}^{\left(1\right)}k_{1}^{\left(2\right)}+\gamma_{1}^{\left(2\right)}k_{-1}^{\left(2\right)}k_{1}^{\left(1\right)}k_{1}^{\left(2\right)}\right.\\
 & \,\,\,\,\,\,\,\,\,\,\,\,\,\,\,\,+\left.\gamma_{-1}^{\left(2\right)}k_{-1}^{\left(2\right)}k_{1}^{\left(1\right)}k_{-2}^{\left(2\right)}+\gamma_{1}^{\left(2\right)}k_{-1}^{\left(2\right)}k_{1}^{\left(1\right)}k_{-2}^{\left(2\right)}+\gamma_{1}^{\left(2\right)}k_{-1}^{\left(2\right)}k_{1}^{\left(1\right)}k_{2}^{\left(2\right)}+\gamma_{-1}^{\left(2\right)}k_{-1}^{\left(1\right)}k_{1}^{\left(2\right)}k_{2}^{\left(2\right)}\right)^{2}
\end{split}
\end{equation}
where

\begin{equation}
\begin{split}D & =\gamma_{-1}^{\left(1\right)^{2}}k_{1}^{\left(1\right)^{2}}k_{-1}^{\left(2\right)^{2}}\gamma_{1}^{\left(2\right)}k_{-2}^{\left(2\right)}\left(k_{-1}^{\left(2\right)}+k_{2}^{\left(2\right)}\right)\left(\gamma_{-1}^{\left(2\right)}k_{-1}^{\left(1\right)}+\gamma_{1}^{\left(2\right)}k_{-1}^{\left(2\right)}\right)\left(\gamma_{-1}^{\left(2\right)}k_{-1}^{\left(1\right)}+\gamma_{1}^{\left(2\right)}k_{-1}^{\left(2\right)}+k_{-1}^{\left(1\right)}k_{-1}^{\left(2\right)}+k_{-1}^{\left(1\right)}k_{2}^{\left(2\right)}\right)\\
 & \,\,\,\,\,\,\times\left[k_{-1}^{\left(2\right)}k_{-2}^{\left(2\right)}/\left(k_{1}^{\left(2\right)}k_{2}^{\left(2\right)}\right)-1\right]
\end{split}
\end{equation}
This expansion of $\frac{p_{+}}{p_{-}}\left(J\right)$, along with
the corresponding exact solution, is plotted in Figure \ref{fig:pfpr-plots}a,
where it is seen that the two match closely for small $J$. Similarly,
for $J\rightarrow0$, we can expand $\frac{p_{+}}{p_{-}}\left(\gamma_{1}^{\left(1\right)}\right)$
about $\gamma_{1}^{\left(1\right)}=\gamma_{1_{\mathrm{c}}}^{\left(1\right)}$---where
$\gamma_{1_{\mathrm{c}}}^{\left(1\right)}$ is the value of $\gamma_{1}^{\left(1\right)}$
that satisfies the hidden detailed balance condition (eq \ref{eq:hdb-enz})---as
$\frac{p_{+}}{p_{-}}\left(\gamma_{1}^{\left(1\right)}\right)\approx\exp\left[\beta w\right]+b\left(\gamma_{1}^{\left(1\right)}-\gamma_{1_{\mathrm{c}}}^{\left(1\right)}\right)^{2}$
(see the Supporting Information for further details). Here, $b=\frac{1}{2}\frac{d^{2}}{d\gamma_{1}^{\left(1\right)^{2}}}\left[\frac{p_{+}}{p_{-}}\left(\gamma_{1}^{\left(1\right)}\right)\right]_{\gamma_{1}^{\left(1\right)}=\gamma_{1_{\mathrm{c}}}^{\left(1\right)}}$,
with $b<0$ ($>0$) for $w>0$ ($<0$). We plot this expansion of
$\frac{p_{+}}{p_{-}}\left(\gamma_{1}^{\left(1\right)}\right)$ in
Figure \ref{fig:pfpr-plots}b, with the corresponding exact solution
shown in Figures \ref{fig:pfpr-plots}b and \ref{fig:pfpr-plots}c;
like before, the two match closely for small $J$. In all panels of
Figure \ref{fig:pfpr-plots}, since $w>0$, it is seen that the bound
$p_{+}/p_{-}\leq\exp\left[\beta w\right]$ is obeyed.

\section{Conclusions}

In this paper, we have employed a stochastic thermodynamics framework
to interpret our recent kinetic analysis \citep{fpt-fluct-thm-arxiv}
of the first-passage time fluctuation theorem, and to generalize it
to a broad class of systems. Based on the generic kinetic model in
Figure \ref{fig:gen-model}a for a cooperative biomolecular machine,
we have demonstrated that, along first-passage trajectories, entropy
production remains constant when the changes in stochastic entropy
and free energy of the machine are equal, which corresponds to zero
net hidden flux through the initial state manifold. Under this condition,
which we have defined quite generally, the first-passage time fluctuation
theorem is restored, with its violation serving as a detectable signature
of hidden detailed balance breaking (which we have also characterized).
Additionally, using the enzymatic model in Figure \ref{fig:enz-model}a,
we have shown that the violation of this relation can be thought of
as a consequence of the breakdown of local detailed balance in the
steps linking $\bar{B}_{1}$ and $\bar{B}_{2}$. In the absence of
hidden current, the fluctuation theorem is restored, and a mesoscopic
local detailed balance condition can be established, which has implications
for the analysis of coarse-grained systems and will be explored further
in a future publication.

\section*{Acknowledgments}

This work was supported by the NSF (Grant No. CHE-1112825) and the
Singapore-MIT Alliance for Research and Technology (SMART). D.E.P.
acknowledges support from the NSF Graduate Research Fellowship Program.

\bibliography{stoch-therm-mach}

\end{document}